\documentstyle[12pt,epsfig]{article}
\newcommand{\be}{\begin{equation}}
\newcommand{\ee}{\end{equation}}
\newcommand{\beq}{\begin{equation}}
\newcommand{\eeq}{\end{equation}}
\newcommand{\ba}{\begin{eqnarray}}
\newcommand{\ea}{\end{eqnarray}}
\newcommand{\bea}{\begin{eqnarray}}
\newcommand{\eea}{\end{eqnarray}}

\begin{document}
\baselineskip=15.5pt \pagestyle{plain} \setcounter{page}{1}
\begin{titlepage}

\leftline{OUTP-10 15 P}

\vskip 0.8cm

\begin{center}

{\LARGE Towards 't Hooft parameter corrections to charge transport
in strongly-coupled plasma} \vskip .3cm

\vskip 1.cm

{\large {Babiker Hassanain\footnote{\tt babiker@thphys.ox.ac.uk}$
^{, \, a}$ and Martin Schvellinger\footnote{\tt
martin@fisica.unlp.edu.ar}$^{, \, b}$} }

\vskip 1.cm

{\it $^a$ The Rudolf Peierls Centre for Theoretical Physics, \\
Department of Physics, University of Oxford. \\ 1 Keble Road,
Oxford, OX1 3NP, UK. \\
$\&$ Christ Church College, Oxford, OX1 1DP, UK.} \\

\vskip 0.5 cm

{\it $^b$ IFLP-CCT-La Plata, CONICET and \\
Departamento  de F\'{\i}sica, Universidad Nacional de La Plata.
\\ Calle 49 y 115, C.C. 67, (1900) La Plata,  \\ Buenos Aires,
Argentina.} \\

\vspace{1.cm}

{\bf Abstract}

\end{center}

We study $R$-charge transport in a wide class of strongly-coupled
supersymmetric plasmas at finite temperature with 't Hooft coupling
corrections. To achieve this, we use the gauge/string duality and
include the full set of ${\cal{O}}(\alpha'^3)$ corrections to the
supergravity backgrounds given at zeroth order by the direct product
of the AdS$_5$-Schwarzschild black hole with a five-dimensional
compact Einstein manifold. On general grounds, the reduction leads
to a large number of higher derivative operators, which we reduce
using the symmetries of the solution. We are left with a universal
set of operators whose coefficients can in principle be fixed by
carrying out an explicit compactification. We apply our results to
the computation of the $R$-charge conductivity of the supersymmetric
plasma at finite yet strong coupling.

\noindent

\end{titlepage}

\newpage


\vfill

\section{Introduction}

The past few years have seen increasing interest in the properties
of the deconfined quark-gluon plasma (QGP), obtained as a result of
the collision of heavy nuclei. Observations at the Relativistic
Heavy Ion Collider (RHIC) imply that once this plasma is produced at
temperatures of order a hundred MeV, it behaves like an ideal fluid.
We refer the reader to several review articles discussing the
phenomenology of the QGP
\cite{Shuryak:2003xe,Gyulassy:2004zy,Muller:2007rs,CasalderreySolana:2007zz,
Shuryak:2008eq,Heinz:2008tv,Iancu:2008sp}. Such experimental
motivation necessitates an understanding of the hydrodynamic
properties of the QGP from the theoretical side, and an excellent tool
to use for this investigation is the AdS/CFT correspondence
\cite{Maldacena:1997re,Gubser:1998bc,Witten:1998qj,Aharony:1999ti}.
The latter conjectures a duality between a class of
highly symmetric strongly-coupled quantum field theories and strings
propagating in certain gravity backgrounds. The reason why the
AdS/CFT correspondence is a good candidate to approach the QGP is
that the latter is strongly-coupled in the relevant regime of
temperature. An immediate focus would be to use the AdS/CFT
correspondence to compute the transport coefficients of
highly-symmetric plasmas in the hydrodynamic regime, defined as the
regime in which the perturbations of the plasma have a momentum much
smaller than the temperature. The important observables, which
hopefully can be used to compare with experiment, are the usual
transport coefficients of fluid dynamics entering the Navier-Stokes
equation. The recipe is to calculate the retarded two-point
functions of conserved currents of the theory at thermal
equilibrium, following the rules established in
\cite{Son:2002sd,Policastro:2002se}. Using these rules, and working
in the gravitational holographic dual model, one is able to obtain
the transport coefficients of both mass (energy-momentum) and
charge, extracting quantities such as viscosity
\cite{Policastro:2001yc,Kovtun:2003wp,Kovtun:2004de} and the
$R$-charge conductivity
\cite{Policastro:2002se,Kovtun:2003wp,CaronHuot:2006te}.

Now we must bear in mind that the results obtained in these
references, and in fact in all of the literature utilizing the
holographic duality, apply to theories with a large number of
degrees of freedom (large-$N$), and with a certain degree of
supersymmetry and/or conformal invariance. These are of course the
limitations of the AdS/CFT correspondence, and so a direct
comparison to experimental observations is difficult. It is
nonetheless important to understand the strong-coupling regime of
these highly-symmetric theories fully, as they share many features
with QCD. In fact, the shear viscosity results for ${\mathcal{N}}=4$
SYM theory \cite{Policastro:2001yc,Kovtun:2003wp,Kovtun:2004de}
obtained via the AdS/CFT correspondence are very close to those
measured for the QGP \cite{Shuryak:2003xe}. Moreover, the AdS/CFT
correspondence allows us to identify universal properties of the
hydrodynamic coefficients \cite{Buchel:2008ae}, which would be very
useful if QCD were to be shown to be within the class of theories in
which this universality is operative. In addition, one may improve
the approach to QCD by incorporating some of its essential aspects
in the gravity dual. This includes adding flavour fields in the
fundamental representation into the gravity dual
\cite{Myers:2007we}, for instance using D3-D7 brane systems as in
the Karch and Katz model \cite{Karch:2002sh}. Another important
direction to pursue is to go to finite coupling, by including
corrections to the infinite 't Hooft parameter results. Within the
context of the AdS/CFT correspondence, this is achieved by adding
string-theoretic higher-curvature corrections to the gravitational
background. This is the premise of the present article.

There is by now a large volume of literature on finite-coupling
corrections to the transport properties of plasma in the
hydrodynamic regime
\cite{Buchel:2008ae,Buchel:2004di,Benincasa:2005qc,Buchel:2008sh,Myers:2008yi,
Ritz:2008kh,Myers:2009ij,Cremonini:2009sy}. Such transport
properties include the shear viscosity and the mass-density
diffusion constants, both of which can be obtained by studying
tensor fluctuations of the supergravity metric with higher-curvature
corrections. On the other hand, the vector fluctuations of the
metric yield quantities such as the $R$-charge diffusion and
conductivity. The finite coupling corrections to the latter
quantities have been considered so far only for the cases where the
additional curvature terms have been of mass-dimension four and six.
In type IIB string theory, the stringy corrections made up of the
metric and the Ramond-Ramond five-form field strength are known
explicitly, and are found to yield dimension eight operators. In
this paper, we analyze the effect of these dimension-eight operators
on the vector fluctuations of the supergravity metric. We should
mention that in a recent paper we have studied the effect of the
full ${\cal {O}}(\alpha'^3)$ corrections from type IIB string
theory, including those derived from the Ramond-Ramond five-form
field strength, on the retarded current-current correlators at the
high energy regime, where the plasma is probed at distances smaller
than the inverse of the temperature \cite{Hassanain:2009xw}.

Let us describe the general idea of the computation we carry out
firstly and summarize our results. The type IIB string theory action
with leading order ${\mathcal{O}}(\alpha'^3)$ corrections, is given
by the usual two-derivative minimal Lagrangian with certain
eight-derivative corrections added. The schematic form of the
${\mathcal{O}}(\alpha'^3)$ corrections is $Weyl^4+ Weyl^3
{\mathcal{T}}+Weyl^2 {\mathcal{T}}^2$, where $Weyl$ is the
ten-dimensional Weyl tensor and ${\mathcal{T}}$ is based on the
five-form field strength $F_5$ of type IIB string theory. A general
solution to this complicated Lagrangian is a {\emph{warped}} product
of a deformed AdS-Schwarzschild black hole with a five-dimensional
Einstein manifold $M_5$, which for instance can be a Sasaki-Einstein
manifold \cite{Buchel:2006gb}. This is dual to a supersymmetric
conformal field theory (SCFT), with an $R$-symmetry group that
contains at least one $U(1)$ factor.

We want to examine the $R$-charge correlators in this dual theory,
focussing on the diagonal $U(1)$ which is dual to the graviphoton.
Note that a weak gauging of this $U(1)$ on the field theory side,
with a small gauge coupling in the manner of
\cite{CaronHuot:2006te}, allows us to interpret our results as
pertaining to an embedding of $U(1)_{em}$ in the $R$-symmetry group
of the field theory. This obviously gives our computations below
added significance, because we may view them as an investigation
into the finite 't Hooft coupling corrections to electromagnetic
charge transport in a wide class of theories. The fields dual to the
field-theory $U(1)$ are vectorial perturbations of the metric and
the five-form field strength. We thus must perturb the background
supergravity solution in the vectorial (mixed) directions
$A_\mu=G_{\mu a}$, plug the perturbed supergravity solution into the
corrected Lagrangian, and integrate out the $M_5$ directions $a$ to
obtain a five-dimensional Lagrangian for the gauge fields $A_\mu$.
At infinite 't Hooft coupling, one simply uses the minimal type IIB
supergravity action, and obtains the Einstein-Maxwell Lagrangian in
AdS-Schwarzschild black hole, {\it i.e.} five-dimensional gravity
(with a cosmological constant) coupled to a $U(1)$ graviphoton. The
computation at finite coupling is much more complicated, as one must
take into account all of the dimension eight string-theory operators
when performing the compactification. To achieve this, we focus our
attention on the construction of the full (complete) set of
five-dimensional operators that can be induced by the
ten-dimensional operators $Weyl^4+ Weyl^3 {\mathcal{T}}+Weyl^2
{\mathcal{T}}^2$. We find 26 such operators, and they have the
schematic form $C^2 F^2$, $C^2 (\nabla F)^2$ and $(\nabla F)^2$,
where $F$ is the field-strength of the $U(1)$ boson. Observe that
these operators will be present in any type IIB string theory dual
model, giving our computation an added incentive. The numerical
values of the coefficients of the operators {\emph{may}} be
dependent on the explicit type of internal manifold, but the hope is
that some of them are universal ({\it i.e.} independent of the
internal manifold $M_5$).

Having obtained the complete five-dimensional Lagrangian with
arbitrary coefficients, we apply our results to obtain the equations
of motion of the transverse gauge field $A_x$, whose solution we
require to compute the $R$-charge conductivity. This field decouples
from the other perturbations, as we show explicitly. We solve the
equations of motion of this field in the hydrodynamic regime as a
series in the momentum, requiring ingoing boundary conditions at the
horizon. We find that the frequency of the waves at the horizon is
unchanged with respect to the infinite 't Hooft coupling results,
for any gauge-invariant set of operators: our results therefore
strongly suggest that any gauge invariant Lagrangian for the vector
perturbations yields an equation of motion with the same singularity
structure and indices at the horizon. We then obtain a general
expression for the leading 't Hooft coupling corrections to the
$R$-charge conductivity.

We view our results as a step towards a better understanding of
charge-transport in strongly-coupled gauge theory plasmas for a
range of theories. We emphasize that the set of operators enumerated
in this work are present in any type IIB string theory holographic
dual model, because the order ${\mathcal{O}}(\alpha'^3)$
ten-dimensional terms are present regardless of the details of the
dual theory ({\it e.g.} whether it has flavour branes or not). Thus,
obtaining this complete set of operators in this setting is of
intrinsic value even if the exact coefficients of the operators in
five dimensions are unknown. The hope is that some (or many) of
these coefficients will be universal, as we speculate in later
sections of the paper.

\section{The corrected background}

Let us define the premise of the paper more carefully: we are
interested in the retarded correlators of the vector currents
associated with a gauged $U(1)$ subgroup of the global $R$-symmetry
group of a SCFT plasma. For example, the conductivity is extracted
from the retarded current-current commutator
\be\label{Jxcorrelator}
R_{\mu\nu}(q) = i \int d^4x \, e^{-i q \cdot x} \, \Theta(x_0) \,
<[J_\mu(x), J_\nu(0)]> \, ,
\ee
where $\Theta(x_0)$ is the Heaviside function, while $J_\mu(x)$ is
the conserved current associated with the gauged $U(1)$ subgroup
mentioned above. The expectation value is understood as a thermal
average over the statistical ensemble of the SYM plasma at
temperature $T$. Let us first consider the string theory holographic
dual description of this theory at infinite 't Hooft coupling. This
is a solution of type IIB supergravity with only the leading
curvature terms, namely the Einstein-Hilbert action coupled to the
dilaton and the Ramond-Ramond five-form field strength:
\be
S_{10}=\frac{1}{2 \kappa_{10}^2}\int \, d^{10}x\,
\sqrt{-G}\left[R_{10}-\frac{1}{2}\left(\partial\phi
\right)^2-\frac{1}{4.5!}\left(F_5\right)^2 \right] \,
.\label{action-10D}
\ee
The solution to the equations of motion of this Lagrangian sourced
by $N$ D3-branes at finite temperature is given by the direct
product of an AdS-Schwarzschild black hole with a compact Einstein
manifold $M_5$. The five-form field strength is given as the sum of
the volume forms on the two manifolds,
\be
F^{(0)}_5 = -\frac{4}{R} (1+\ast) \epsilon_5 \, ,
\label{F5-definition}
\ee
where the superscript $^{(0)}$ indicates that this is a pure
supergravity solution, {\it i.e.} with no stringy corrections. Its
total flux through the compact manifold gives $N$ units. The current
operator $J_\mu(x)$ is dual to the $s$-wave mode of the vectorial
fluctuation about this background. In order to obtain the correct
AdS-Schwarzschild Lagrangian for the vectorial perturbation, one
must construct a consistent perturbed {\it ansatz} for both the
metric and the five-form field strength, such that a proper $U(1)$
subgroup of the $R$-symmetry group is obtained (see
\cite{Argurio:1998cp,Cvetic:1999xp,Chamblin,Tran} for the $S^5$
solution, and \cite{Buchel:2006gb} for the five-dimensional
Sasaki-Einstein solution). The result of inserting the consistent
perturbation {\it ansatz} into the minimal type IIB supergravity
action is the minimal $U(1)$ gauge field kinetic term in the
AdS-Schwarzschild black hole. Therefore, by studying the bulk
solutions of the Maxwell equations in the AdS-Schwarzschild black
hole with certain boundary conditions, we can obtain the retarded
correlation functions \cite{Son:2002sd,
Policastro:2002se,CaronHuot:2006te} of the operator $J_\mu(x)$. Our
aim in this section is to describe this procedure for the
$\alpha'^3${\emph{-corrected}} type IIB supergravity action, which
contains dimension-eight higher curvature operators. These
higher-curvature corrections on the supergravity side correspond to
finite-coupling corrections in the field theory, hence our interest
in their effect. Essentially, for any given field-theoretic
observable $\mathcal{O}$, we can write a series
${\mathcal{O}}_0+{\mathcal{O}}_1/\lambda^{n_1}+\cdots$, where
$\lambda$ is the 't Hooft coupling, and $n_1$ is a positive number
which indicates that the lowest order correction to the result at
infinite coupling ${\mathcal{O}}_0$ need not begin at order one. We
are interested in the case where ${\mathcal{O}}_0$ is the two-point
current correlator, and we must thus determine the effect of the
string-theory corrections on the vectorial perturbations of the
metric.

The inclusion of higher-derivative corrections to the supergravity
must take place at the level of the ten-dimensional action, through
the evaluation of stringy corrections to Eq.(\ref{action-10D}). The
leading corrections were found to begin at
${\mathcal{O}}(\alpha'^3)$. These corrections were found to have no
effect on the metric at zero temperature \cite{Banks:1998nr},
verifying certain non-renormalization theorems of CFT correlators.
At finite temperature \cite{Gubser:1998nz,Pawelczyk:1998pb}, much of
the work focussed on the corrections to the thermodynamics of the
black hole. The corrections were then revisited in references
\cite{deHaro:2002vk,deHaro:2003zd,Peeters:2003pv}, where the
computation of the $\alpha'$-corrected metric was improved and
attempts were made to address the issue of the completeness of the
corrections at leading order in $\alpha'$. Much of the interest of
the community has focussed on the effect of higher curvature
corrections on the spin-2 sector of the fluctuations
\cite{Buchel:2004di,Buchel:2008sh,Hofman:2009ug,Sinha:2009ev}, as
these determine the viscosity and mass-diffusion constants of the
plasma. In \cite{Paulos:2008tn,Myers:2008yi} the higher curvature
corrections to the dual of  ${\cal {N}}=4$ SYM were parsed
thoroughly to determine how they affect the metric. Crucially, the
corrections to the metric were found to be universal, {\it i.e.}
independent of the internal manifold for a wide class of internal
manifolds, of which Sasaki-Einstein is a member.

For the case of the vectorial fluctuations of the background, there
are two distinct parts to the calculation: the first part consists
of obtaining the minimal gauge-field kinetic term using new
perturbed {\emph{and}} corrected metric and five-form field strength
{\it ans${\ddot{a}}$tze}. The second part of the computation
consists of obtaining the corrections to the gauge field Lagrangian
coming directly from the higher-derivative operators. The reason why
these two steps are distinct is that the first step will require
insertion of the corrected perturbation {\it ans${\ddot{a}}$tze}
into the minimal ten-dimensional supergravity two-derivative part
Eq.(\ref{action-10D}). The second step requires insertion of the
{\it{uncorrected}} perturbation {\it ans${\ddot{a}}$tze} into the
higher-curvature terms in ten dimensions, for consistency in the
$\alpha'$ expansion.

Below we choose to use a specific manifold, the $S^5$ manifold, as
an illustration of the methodology, but we emphasize to the reader
that the discussion below applies directly without modification to
any five-dimensional compact Einstein manifold. The only restriction
comes from the requirement that in the zeroth-order background
supergravity solution the only non-trivial fields are the metric and
the five-form field strength. Our discussion thus applies to all
supergravity backgrounds with an internal component satisfying these
requirements, and these internal manifolds happen to be Einstein.
Thus, we could have used the {\it ans${\ddot{a}}$tze} espoused in
\cite{Buchel:2006gb} to arrive at the same conclusions. We only use
the five-sphere in what follows for simplicity and familiarity.

We begin by examining the corrected metric and $F_5$ solutions, then
proposing {\it ans${\ddot{a}}$tze} for the perturbations that may be
inserted into  Eq.(\ref{action-10D}) to obtain the minimal gauge
kinetic term. The corrections to the ten-dimensional type IIB action
are given by \cite{Myers:2008yi}
\be\label{10DWeyl}
S_{10}^{\alpha'}=\frac{R^6}{2 \kappa_{10}^2}\int \, d^{10}x\,
\sqrt{-G}\left[ \, \gamma e^{-\frac{3}{2}\phi}W_4 + \cdots\right] \, ,
\ee
where $\gamma$ encodes the dependence on the 't Hooft coupling
$\lambda$ through the definition $\gamma \equiv \frac{1}{8} \,
\xi(3) \, (\alpha'/R^2)^{3}$, with $R^4 = 4 \pi g_s N \alpha'^2$.
Setting $\lambda = g_{YM}^2 N \equiv 4 \pi g_s N$, we get
$\gamma =   \frac{1}{8} \, \xi(3) \, \frac{1}{\lambda^{3/2}}$.

The $W_4$ term is a dimension-eight operator, and is given by
\be
W_4=C^{hmnk} \, C_{pmnq} \, C_h^{\,\,\,rsp} \, C^{q}_{\,\,\,rsk} +
\frac{1}{2} \, C^{hkmn} \, C_{pqmn} \, C_h^{\,\,\, rsp} \,
C^q_{\,\,\, rsk} \, ,
\ee
where $C^{q}_{\,\,\, rsk}$ is the Weyl tensor. The dots in
Eq.(\ref{10DWeyl}) denote extra corrections containing contractions
of the five-form field strength $F_5$, which we can schematically
write as $\gamma
(C^3{\mathcal{T}}+C^2{\mathcal{T}}^2+C{\mathcal{T}}^3+{\mathcal{T}}^4)$,
where $C$ is the Weyl tensor and ${\mathcal{T}}$ is a tensor found
in \cite{Myers:2008yi} and composed of certain combinations of
$F_5$. The authors of \cite{Myers:2008yi} showed that the metric
itself is only corrected by $W_4$, essentially due to the vanishing
of the tensor ${\mathcal{T}}$ on the uncorrected supergravity
solution. This conclusion was found to be independent of the
internal manifold, as long as the only non-trivial fields in the
zeroth-order supergravity background are the metric and the
five-form field strength. As we mentioned this holds for any compact
Einstein manifold. Hence, all Sasaki-Einstein manifolds $L_{pqr}$,
as well as $Y^{p,q}$ and $T^{1,1}$ which are special cases of them,
are covered by what follows.

After taking into account the contribution of $W_4$ to
the Einstein equations, one finds the corrected metric
\cite{Gubser:1998nz,Pawelczyk:1998pb,deHaro:2003zd}
\be
ds^2 = \left(\frac{r_0}{R}\right)^2\frac{1}{u} \, \left(-f(u) \,
K^2(u) \, dt^2 + d\vec{x}^2\right) + \frac{R^2}{4 u^2 f(u)} \,
P^2(u) \, du^2 + R^2 L^2(u) \, d\Omega_5^2 \, ,\label{proper-metric}
\ee
where $f(u)=1-u^2$ and $R$ is the radius of the AdS$_5$. In these
coordinates the AdS-boundary is at $u=0$ while the black hole
horizon is at $u=1$. We denote the AdS$_5$ coordinates by the
indices $m$, where $m= \{(\mu=0, 1, 2, 3),5\}$, where
\be
K(u) = \exp{[\gamma \, (a(u) + 4b(u))]} \, ,  \quad P(u) =
\exp{[\gamma \, b(u)]} \, , \quad L(u) =  \exp{[\gamma \, c(u)]} \,
,
\ee
and
\ba
a(u) &=& -\frac{1625}{8} \, u^2 - 175 \, u^4 + \frac{10005}{16} \,
u^6 \, , \quad b(u) = \frac{325}{8} \, u^2 + \frac{1075}{32} \, u^4
- \frac{4835}{32} \, u^6 \, , \nonumber \\
c(u) &=& \frac{15}{32} \, (1+u^2) \, u^4 \, , \quad \quad
\textrm{with} \quad \quad r_0 = \frac{\pi T R^2}{(1+\frac{265}{16}
\gamma)} \, ,
\ea
where $T$ is the physical temperature of the plasma. Having obtained
the corrected metric, the next step is to deduce the appropriate
perturbation {\it ans${\ddot{a}}$tze}. This is in fact where the
complications of the problem appear: the vector perturbation enters
into both the perturbed metric and the perturbed $F_5$ solution.
This is distinct from the case where one considers tensor
perturbations of the background, which are relevant for viscosity
computations, because they only enter into the metric {\it ansatz},
not into the $F_5$ {\it ansatz}, making the computations far
simpler.

Let us first consider how we would obtain the minimal
(two-derivative) kinetic term for the gauge fields. We must insert
our corrected {\it ansatz} into the two-derivative supergravity
action Eq.(\ref{action-10D}). The metric {\it ansatz} we use is as
follows, with obvious substitutions, where we have imposed that the
internal metric is the five-sphere (for ease of demonstration)
\be\label{metric-ansatz}
ds^2=g_{mn} \,dx^m \, dx^n + R^2 L(u)^2 \, \sum_{i=1}^3 \left[
d\mu_i^2+\mu_i^2(d\phi_i+\frac{2}{\sqrt{3}}A_\mu dx^\mu)^2 \right]
\, ,
\ee
where the $\mu_i$ are the direction cosines for the sphere, as usual.

As for the {\it ansatz} for $F_5=G_5+\ast G_5$ we use the fact that
we are only interested in the terms which are quadratic in the
gauge-field perturbations. Thus we use the following {\it ansatz}
\be
G_5 = -\frac{4}{R} \overline{\epsilon}_5 + \frac{R^3
L(u)^3}{\sqrt{3}} \, \left( \sum_{i=1}^3 d\mu_i^2 \wedge d\phi_i
\right) \wedge \overline{\ast} F_2 \, , \label{F5ansatzCorrected}
\ee
where $F_2 = dA$ is the Abelian field strength and
$\overline{\epsilon}_5$ is the deformation of the volume form of the
five-dimensional metric of the AdS-Schwarzschild black hole
\footnote{Note that we are not interested in the part of $G_5$ which
does not contain the vector perturbations. This part only
contributes to the potential of the metric, and is thus accounted
for by the use of the corrected metric in the computation.}. The
Hodge dual $\ast$ is taken with respect to the ten-dimensional
metric, while $\overline{\ast}$ denotes the Hodge dual with respect
to the five-dimensional metric piece of the black hole. Note that we
have not dwelled on the details of the {\it ansatz} because the main
point of the paper is that the operators derived below are in fact
independent of the internal manifold in form. The only dependence
comes in through their coefficients. Inserting these {\it ans\"atze}
into Eq.(\ref{action-10D}), and discarding all the higher (massive)
Kaluza-Klein harmonics of the five-sphere, we get the following
action for the zero-mode Abelian gauge field $A_m$:
\be
S = -\frac{\tilde{N}^2}{64 \pi^2 R} \int d^4x \, du \, \sqrt{-g} \,
L^7(u) \, g^{mp} \, g^{nq} \, F_{mn} \, F_{pq} \, , \label{Fsquared}
\ee
where the Abelian field strength is $F_{mn}=\partial_m A_n -
\partial_n A_m$, the partial derivatives are $\partial_m =
\partial/\partial x^m$, while $x^m=(t, \vec{x}, u)$, with $t$ and
$\vec{x}=(x_1, x_2, x_3)$ being the Minkowski coordinates, and $g
\equiv \textrm{det} (g_{mn})$, where the latter is the metric of
AdS-Schwarzschild black hole piece of the corrected metric. The
dependence on the dimensionless factor $L(u)$ is acquired by the
proper reduction from ten dimensions \cite{Kovtun:2003wp}, and
ultimately arises as a consequence of the non-factorisability of the
corrected metric \cite{Pawelczyk:1998pb}. This factor is independent
of the internal metric $M_5$. Note also that the volume of the
internal manifold has been absorbed into the definition of the
factor $\tilde{N}$.

We have thus completed  the first step in our programme, that of
obtaining the minimal gauge kinetic term from the two-derivative
supergravity action. The next step is to obtain the effect of the
eight-derivative corrections of Eq.(\ref{10DWeyl}). Concretely, we
must determine the five-dimensional operators that arise once the
perturbed metric and five-form field strength {\it
ans${\ddot{a}}$tze} of equations (\ref{F5ansatzCorrected}) and
(\ref{metric-ansatz}) are inserted into Eq.(\ref{10DWeyl}).
Crucially, we are able to use the uncorrected {\it
ans${\ddot{a}}$tze} in this step, because using the corrected ones
results in terms of even higher order in $\gamma$. The uncorrected
{\it ans${\ddot{a}}$tze} are derived from those of equations
(\ref{F5ansatzCorrected}) and (\ref{metric-ansatz}) by taking
$L(u),K(u),P(u)\to 1$ and $\overline{\epsilon}_5 \to \epsilon_5$.
Our philosophy will be to consider the structure of the $C^3 {\cal
{T}}$ and $C^2 {\cal {T}}^2$ terms, and use the symmetries of the
various tensors to deduce the most general set of five-dimensional
operators that can be obtained via the compactification. We describe
how this is achieved in the next section.

\section{Operator enumeration}

The ten-dimensional corrections in totality are schematically given by
\begin{equation}
C^4 + C^3 {\cal {T}} + C^2 {\cal {T}}^2 + C {\cal {T}}^3 + {\cal {T}}^4 \,, \label{TensorT}
\end{equation}
where ${\cal {T}}$ is given by
\begin{equation}
{\cal {T}}_{abcdef}= i\nabla_a
F^{+}_{bcdef}+\frac{1}{16}\left(F^{+}_{abcmn}F^{+}_{def}{}^{mn}-
F^{+}_{abfmn}F^{+}_{dec}{}^{mn}\right) \, ,
\end{equation}
where the RHS must be antisymmetrized in $[a,b,c]$ and $[d,e,f]$ and
symmetrized with respect to interchange of $abc \leftrightarrow def$
\cite{Paulos:2008tn} and we have defined the tensor
\be
F^{+} = \frac{1}{2} (1+\ast) F_5 \, .
\ee
The perturbed {\it ansatz} for the five-form field strength contains
only one power of the gauge field strength. Therefore, we can write
${\cal {T}} = {\cal {T}}_0 + {\cal {T}}_1 + {\cal {T}}_2$, with each
subscript denoting the number of powers of the gauge field contained
in the tensor. The tensor ${\cal {T}}_0$ is zero for all
supergravity backgrounds given by a direct product of an
AdS$_5$-Schwarzschild black hole with a five-dimensional compact
Einstein manifold, provided that the five-form field strength
$F^{(0)}_5$ can be written as in Eq.(\ref{F5-definition})
\cite{Myers:2008yi}. Therefore, splitting the ten-dimensional Weyl
tensor in a similar fashion, the corrections can be schematically
written as
\begin{equation}
C_0^2 C_1^2 + C_0^3 C_2+C_0^2 C_1 {\cal {T}}_1 + C_0^3 {\cal {T}}_2 + C_0^2 {\cal {T}}_1^2   \,.
\end{equation}
Now, let us study the term $C_0^3 {\cal {T}}_2$. As we are
considering a direct product space, the Weyl tensor factorizes into
its AdS-Schwarzschild black hole and $M_5$ factors. Hence, $C_0^3$
must reside entirely in one of the two factors. If it resides in the
AdS-Schwarzschild black hole factor, then the only five-dimensional
operators that can result from this are of the form
\begin{equation}
\tilde{C}^3 F^2 \, ,
\end{equation}
where $\tilde{C}$ denotes the Weyl tensor of five-dimensional
AdS-Schwarzschild black hole and $F$ is the $U(1)$ field strength
tensor. We have checked explicitly that, for any given $M_5$, the
non-zero entries of ${\cal {T}}_2$ correspond to zero entries of
$\tilde{C}^3$, and so this operator vanishes generally, and
operators of the form $\tilde{C}^3 F^2$ are therefore not induced.
The same argument can be constructed for the contribution of $C_0^3
C_2$. However, the tensor $C_0^3$ may reside in the $M_5$ factor, in
which case the induced operators in five dimensions are of the form
$F^2$. Of course, there is only one such operator, proportional to
the kinetic term.

Let us now focus on the terms given by $C_0^2 {\cal {T}}_1^2$ and
$C_0^2 C_1^2$. Again, $C_0^2$ factorizes; if it resides in the
AdS-Schwarzschild black hole factor, then we must obtain operators
of the form
\begin{equation}
\tilde{C}^2 F^2 \quad \textrm{and}  \quad \tilde{C}^2 (\nabla F)^2 \, .
\end{equation}
If $C_0^2$ resides in the $M_5$ factor, then we must obtain all
operators of the form
\begin{equation}
F^2 \quad \textrm{and} \quad (\nabla F)^2 \, .
\end{equation}
A bit of thought should then convince the reader that the same
considerations apply to $C_0^2 C_1 {\cal {T}}_1$ because in this way
of thinking $C_1$ is entirely equivalent to ${\cal {T}}_1$ in that
it also contains only one power of the gauge field. Therefore, the
problem reduces to finding the set of independent monomials
comprising all contractions of two Weyl tensors and two $\nabla F$,
as well as two Weyl tensors and two $F$. In addition, we require a
set of monomials to represent all contractions of two $\nabla F$,
and all contractions of two powers of $F$.

This can all be very quickly computed by Cadabra, the program
developed by Kasper Peeters. We find 26 such operators in total, and
we list them here. One of them may be eliminated on shell (it does
not contribute to the on-shell action), and so the final set
contains 25 operators in total. Note that these comprise a full
linearly-independent set up to the use of dimension dependent
identities that are similar in nature to the Schouten identities of
\cite{Buchel:2008ae}. We expect these identities to reduce the set
by four, but we have not undertaken the reduction in what follows.
We would like to stress that this result is indeed a massive
reduction compared with the very large number of general
five-dimensional operators which are possibly induced by the
five-dimensional reduction of the ten-dimensional operators of
Eq.(\ref{TensorT}) upon a general compact Einstein manifold. Just to
give an idea of this consider for instance that an operator like
$C^2 (\nabla F)^2$ leads to 720 distinct operators induced from the
permutations of the operator $(\nabla F \nabla F)_{abcdef}$ before
any symmetry operations are taken into account.

The full set of 26 five-dimensional scalar operators is given by
\begin{eqnarray}\label{effective-L}
&& (C^4+C^3{\mathcal{T}}+C^2{\mathcal{T}}^2+C{\mathcal{T}}^3)|_{5d} = \nonumber \\
&&a_1 C_{abcd} C^{abcd} \nabla^e F_{ef} \nabla^g F^f{}_g
+ a_2 C_{abcd} C^{acbd} \nabla_e F_{fg} \nabla^f F^{eg} + \nonumber \\
&& C_{abc}{}^d C^{acbe} \left[ b_1 \nabla_f F_d{}^f \nabla_g F_e{}^g
+ _2 \nabla_d F_{ef} \nabla_g F^{fg} + b_3 \nabla_d F_{fg} \nabla_e F^{fg}\right]+ \nonumber \\
&& b_4 C_{abc}{}^d C^{abce} \nabla_f F_{dg} \nabla^fF_e{}^g + \nonumber \\
&& C_{a}{}^b{}_c{}^d C^{aecf} \left[ c_1 \nabla_b F_{eg} \nabla^g F_{df}
+ c_2 \nabla_b F_{de} \nabla^g F_{fg} + c_3 \nabla_b F_{dg} \nabla_f F_e{}^{g}
+ c_4 \nabla_b F_{eg} \nabla_d F_f{}^{g} \right] +\nonumber \\
&& C_{a}{}^b{}_c{}^d C^{acef} \left[ c_5 \nabla_b F_{dg} \nabla_e F_{f}{}^g
+ c_6 \nabla_b F_{ef} \nabla_g F_{d}{}^g + c_7 \nabla_b F_{eg} \nabla_f F_d{}^{g} \right] + \nonumber \\
&&  c_8 C_{ab}{}^{cd} C^{abef} \nabla_c F_{eg} \nabla^g F_{df} + \nonumber \\
&& C_{a}{}^{bcd} C^{aefg} \left[ d_1 \nabla_b F_{ce} \nabla_f F_{dg}
+ d_2 \nabla_c F_{be} \nabla_f F_{dg}+d_3 \nabla_b F_{cf} \nabla_g F_{de}
+ d_4 \nabla_c F_{bd} \nabla_f F_{eg} \right] +\nonumber \\
&& e_1 C_{abcd} C^{acbd} F_{ef} F^{ef} + f_1 C_{abc}{}^d C^{acbe} F_{df} F_e{}^f + \nonumber \\
&&  g_1 C_{a}{}^b{}_c{}^d C^{aecf} F_{be} F_{df}
+ C_{a}{}^b{}_c{}^d C^{acef} \left[ g_2 F_{bd} F_{ef} + g_3 F_{be} F_{df} \right] + \nonumber \\
&& h_1 F_{ab} F^{ab}+ h_2 \nabla_a F_{bc} \nabla^b F^{ac} + h_3
\nabla^b F_{bc} \nabla^aF_a{}^c \, .
\end{eqnarray}
We remind the reader that this does not mean that all of these will be induced by
the compactification: in all probability only a handful of them will be induced, but
the statement we can definitively make is that the operators listed here comprise the
most complete allowed set. Note that the final operator, with
coefficient $h_3$, can be eliminated on-shell, and we do this in
what follows. Note also that for the sphere the coefficients
$h_i=0$, because the sphere is Weyl flat. This is not necessarily
the case for other manifolds, unless there is a miraculous
cancellation at work. Perhaps this points the way to a
non-universal behaviour in this particular sector of the dual field theory, and we shall have more to
say on this later.

\section{The Lagrangian for the transverse modes}

As an application of the above, we will consider the two-point
correlators of the $R$-symmetry current $J_x$. The dual field on the
supergravity side is the gauge field in the $x$-direction $A_x$.
Two-point functions of $J_x$ are useful for a range of physical
phenomena, including the conductivity and diffusion constant of the
EM charge, and the computation of photoemission spectra by
gauge-theory plasma \cite{CaronHuot:2006te}. Specifically, we will
compute the leading 't Hooft coupling corrections to the
conductivity associated with the $U(1)$ $R$-charge. The conductivity
is given by the following relationship:
\be
\sigma=-i \lim_{\omega\to0}\frac{1}{2 T}R_{xx}(\omega,q=0) \, ,
\ee
where we have used Eq.(\ref{Jxcorrelator}) and the Kubo formula from
reference \cite{CaronHuot:2006te}. Because we are interested only in
the size of the corrections to this quantity, we will measure our
conductivity below in terms of the uncorrected conductivity, so that
our result will have the form $1+\frac{\rho}{\lambda^{3/2}}$, where
$\rho$ is the number we shall compute below.

As we saw in the last section, the eight-derivative
${\mathcal{O}}(\alpha'^3)$ corrections introduce a multitude of
higher-derivative operators upon compactification on the $M_5$,
and we must take account of them properly to solve the
equation of motion within perturbation theory. The situation is
entirely analogous to that studied by Buchel, Liu and Starinets in
\cite{Buchel:2004di}. In that paper, the authors were concerned with
the tensor perturbations of the metric, but the logic is the same.
From our effective Lagrangian, we see immediately that $A_x$ decouples
from the other perturbations. Computing the effect of the minimal kinetic term of Eq.(\ref{Fsquared})
and the above general set of operators of Eq.(\ref{effective-L}) yields the following Lagrangian for the transverse mode $A_x$
\ba
S_{\textrm{total}} &=&- \frac{{\tilde {N}}^2 r_0^2}{16 \pi^2 R^4}
\int \frac{d^4k}{(2 \pi)^4} \int_0^1 du \,
\left[\gamma A_W A_k'' A_{-k}+ (B_1+\gamma B_W) A_k' A_{-k}' \right. \nonumber \\
&&\left. + \gamma C_W A_k' A_{-k}+(D_1+\gamma D_W) A_k A_{-k}
+\gamma E_W A_k'' A_{-k}''+\gamma F_W A_k'' A_{-k}' \right]  \, ,
\label{AxAction}
\ea
where we have introduced the following Fourier transform of the field $A_x$
\be
A_{x}(t, \vec{x}, u) = \int \frac{d^4k}{(2 \pi)^4}
\, e^{-i \omega t + i q z} \, A_k(u) \, .
\ee
There are also a number of boundary terms that must be included for
this higher-derivative Lagrangian to make sense, and this is
discussed in detail in \cite{Buchel:2004di,Hassanain:2009xw}.
The coefficients $B_1$ and $D_1$ arise directly from the minimal kinetic
term $F^2$. The subscript $W$ indicates that the particular
coefficient comes directly from the eight-derivative corrections,
and the functions $A_W \to F_W$ are listed in the appendix.
Moreover, $B_1$ and $D_1$ contain some $\gamma$-dependence, but they
are non-vanishing in the $\gamma \to 0$ limit, while every other
coefficient vanishes in that limit. The equation of motion is given
by
\be
A_{x}''+p_1 A_{x}'+ p_0 A_{x} \, = \gamma \frac{1}{2 f(u)}V(A_{x})
\, ,
\ee
where
\ba
&&A_W \, A_x''+C_W A_x'+2 \left(\delta D_1+D_W\right) A_x \nonumber \\
&&-\partial_u\left(2 \delta B_1A_x'+2 B_W A_x'+C_W A_x +F_W A_x'' \right) \nonumber \\
&&+\partial_u^2 \left(A_W A_x+2 E_W A_x''+F_W A_x'\right)\, = V(A_{x}) \, ,
\ea
where  $B_1-B_1|_{\gamma\to 0}=\delta B_1$ and $D_1-D_1|_{\gamma\to
0}=\delta D_1$.
First we have the coefficients with no
$\gamma$-dependence $p_0$ and $p_1$, given by
\be
p_0= \frac{\varpi_0^2-f(u)\kappa_0^2}{u f^2(u)} \quad \textrm{and} \quad p_1=\frac{f'(u)}{f(u)} \, ,
\ee
where $\varpi_0=\omega/(2 \pi T)$ and $\kappa_0=q/(2 \pi T)$. For the
coefficients originating from the $F^2$ term in the action of the
gauge field, we obtain
\ba
B_1&=&\frac{K(u)f(u) L^7(u)}{P(u)} \, , \nonumber \\
D_1&=&-K(u)P(u) L^7(u)\left[\frac{\varpi^2-f(u)K^2(u)\kappa^2}{u f(u)K^2(u)} \right] \, ,
\ea
where $\varpi=\omega R^2/(2 r_0)$ and  $\kappa=q R^2/(2 r_0)$. The
terms originating from the higher curvature terms in the action are
listed in the appendix.
At this stage it is convenient to reduce the equation to a
second-order differential equation using a simple trick
\cite{Cremonini:2009sy}. The idea is that $\gamma A_x''=-\gamma
\left(p_1 A_{x}'+ p_0 A_{x}\right)+{\mathcal{O}}(\gamma^2)$. Thus,
we may reduce the entire RHS of the equation of motion to terms
which are first or zeroth order in derivatives. The resulting
equation is given by:
\be
A_{x}''+\left[ p_1-\frac{\gamma}{2 f(u)} \, [\theta_1(u)-p_1
\theta_2(u)] \right] A_{x}' + \left[p_0-\frac{\gamma}{2 f(u)} \,
[\theta_0(u)-p_0 \theta_2(u)] \right] A_{x} \, =
{\mathcal{O}}(\gamma^2) \, ,
\ee
where
\ba
&&\theta_0(u)=2 \, (\delta D_1 + D_W)- C_W' + A_W''
- 4 E_W' \, p_0' + 2 \, E_W \, (p_1 p_0'-p_0'') \nonumber \\
&&\theta_1(u)=2 \, A_W'- 2 \, (\delta B_1+B_W)' + F_W''
- 4 E_W' \, (p_1'+p_0) + 2 E_W \, [p_1(p_1'+p_0) - p_1'' - 2 p_0'] \nonumber \\
&&\theta_2(u)=2 \, A_W - 2 \, (\delta B_1+B_W) + F_W' + 2 E_W''-4
E_W' \, p_1 + 2 E_W \, [p_1^2-2 p_1'-p_0] \, .
\ea
We are now in a position to solve this equation in the hydrodynamic
regime. The first step is to examine the singularity structure of
the equation at the horizon $u=1$. As usual, we change variables to
$x=1-u$, so that the singularity is at $x=0$, then insert the
functional form $A_x=x^\beta$. We obtain the indicial equation:
\be
\beta^2+\left(\frac{\omega}{4 \pi T}\right)^2=0 \,.
\ee
This is of course the same indicial equation that would have been
obtained in the infinite 't Hooft coupling limit. Thus, as long as
the Lagrangian originates from a gauge-invariant series of
operators, then the indicial equation is unchanged. We have made
several checks of this statement, using operators of arbitrary
dimension, containing up to four derivatives of the gauge field.
Hence, the fact that the indicial equation is unchanged is a
consequence of the gauge-invariance in five dimensions, which is in
turn a consequence of the $U(1)$ isometry of the internal manifold $M_5$, and
has nothing to do with supersymmetry. We are aware that this
behaviour is expected
\cite{Buchel:2008sh,Ritz:2008kh,Myers:2009ij,Cremonini:2009sy} and
in fact Buchel mentions it in his paper \cite{Buchel:2008sh},
focussing on scalar and tensor fluctuations of the metric. It would
be very interesting to find a general proof of this statement.

\section{Solving the equations of motion in the hydrodynamic regime}

We now turn to the solution of the equations for $A_x$.
In order to compute the conductivity, we must solve the equation for $A_x$
up to linear order in $\gamma$ and $\omega$, which is of course the
hydrodynamic regime. Guided by our observations of the behaviour at the
horizon, we propose the following form for the solution:
\be
A_x(u)=A_0(u)+\gamma A_1(u)\, = \,\left[1-u \right]^{-\sigma} \left(
\phi_0(u)+\gamma \phi_1(u) \right)\, ,
\ee
where $\sigma=i\omega/(4\pi T)$. We now write
$\phi_{0,1}(u)=h_{0,1}(u)+\sigma g_{0,1}(u)$. We insert this
decomposition into the equation of motion, and realize immediately
that the only relevant terms are $B_{1,W}$ (in the limit $\omega$
and $q \to 0$), $F_W$ and $E_W$, because everything else enters with
at least quadratic powers in $\omega$ and $q$ (see the appendix for
$B_W, E_W,F_W$). This is actually a consequence of gauge-invariance:
any gauge-invariant Lagrangian of the form of Eq.(\ref{AW}) will
have this property. This is {\emph{not}} the case for
tensor-fluctuations of the metric.

First we focus on $\phi_0(u)$. Comparing powers of $\sigma$ in the
equation of motion, we simply obtain
\be
h_0(u)=C \quad \textrm{and} \quad g_0(u)=C \log(1+u)+D \,,
\ee
where we will not fix any of the constants $C,D$ until the very end.
The only requirement at this stage is regularity of all of the
functions at the horizon. We now turn to $A_1=\left[1-u
\right]^{-\sigma}\phi_1(u)$. This function obeys a rather
complicated equation of motion which can be deduced
straightforwardly from the parent equation:
\be
A_{1}''+p_1 A_{1}'+ p_0 A_{1} \, = \frac{1}{2 f(u)}V(A_{0}) \, .
\ee
We obtain the following equation for $h_1(u)$
\ba
&& f(u)h_1''+f'(u)h_1'=E_W'' h_0''+2 E_W'h_0'''+E_W h_0'''' \,    \nonumber \\
&+&\frac{1}{2}\left\{\left(F_W'-2\left(\delta B_1+B_W \right)\right)h_0''
+\left(F_W''-2\left(\delta B_1'+B_W' \right)\right)h_0' \right\}    \, ,
\ea
which solves to $h_1(u)=C_\gamma$, also a constant. We must now
solve for $g_1(u)$. Using the explicit forms for $h_{0,1}(u)$ and
$g_0(u)$, the equation for $g_1(u)$ simplifies to
\ba
&& \partial_u\left(f(u)g_1(u)'+\frac{C_\gamma f(u)}{1-u} \right) \,    \nonumber \\
&=&\frac{1}{2}\partial_u\left\{\left(F_W'-2\left(\delta B_1+B_W
\right)\right)\left[g_0'(u)+\frac{C}{1-u} \right]
\right\}+\partial_u^2\left\{ E_W \left[g_0''(u)+\frac{C}{(1-u)^2}
\right] \right\}.
\ea
Note the appearance of the combination $F_W'-2\left(\delta B_1+B_W
\right)$, which could have been anticipated from the work of \cite{Myers:2009ij}.
The above equation for $g_1(u)$ solves to
\begin{equation}
g_1(u)=\theta \log(1+u)+(C_\gamma-\theta)\log(1-u)+D_\gamma +
\Phi(u) \, ,
\end{equation}
where $\Phi(u)$ is given by the following integrals
\begin{equation}
\Phi(u)= C \int du \frac{1}{f^2(u)}\left(F_W'-2(\delta
B_1+B_W)\right)+4 C \int du \frac{1}{f(u)}\partial_u\left(E_W \,
u/f^2(u)\right) \,.
\end{equation}
Therefore, the function $g_1(u)$ is given by
\begin{eqnarray}
g_1(u)&=&\theta \log(1+u)+(C_\gamma-\theta)\log(1-u)+D_\gamma \nonumber \\
&+& C\left( \left( \frac{185}{4}+2 \tilde{\alpha} \right)u
+\left(\frac{185}{8}+ z\right) \log\left[\frac{1-u}{1+u}\right]\right)
+{\mathcal{O}}(u^2)  \, , \nonumber \\
\end{eqnarray}
where $\tilde{\alpha}$ is composed of the coefficients of the Lagrangian
\begin{eqnarray}
\tilde{\alpha}&=&216 a_2 + 144 b_3 + 192 b_4 + 30 c_1 + 54 c_3 - 12 c_4 + 6 c_5 - 60 c_7 - \nonumber \\
 &&12 c_8 - 12 d_1 - 18 d_2 + 18 d_3 - 36 e_1 - 8 f_1 + 5 g_1 - 2 g_2 - g_3 +
 h_2 \, ,
\end{eqnarray}
and $z$ drops out upon requiring regularity at the
horizon. This yields the following solution to linear order in $u$
\begin{equation}
g_1(u)=\left(C_\gamma +C\left[\frac{185}{4}+ 2 \tilde{\alpha} \right] \right)u +D_\gamma \, .
\end{equation}
We now have the full solution of the equations of motion to linear
order in $\gamma$ and $\sigma$:
\begin{equation}
A_x(u)=\left[1-u \right]^{-\sigma} \left( C+\gamma C_\gamma+\sigma
\left\{D+\gamma D_\gamma+\left( C+\gamma C_\gamma
+\gamma C\left[\frac{185}{4}+2 \tilde{\alpha} \right] \right)u  \right\} \right) \,.
\end{equation}
Letting $\overline{C}=C+\gamma C_\gamma$ and $\overline{D}=D+\gamma
D_\gamma$, we then have that, to linear order in $\gamma$
\begin{equation}
A_x(u)=\left[1-u \right]^{-\sigma} \left(\overline{C}+\sigma
\left\{\overline{D}+\overline{C}\left(1 +\gamma \left[\frac{185}{4}
+2\tilde{\alpha} \right] \right)u  \right\} \right) \,.
\end{equation}
If we call the boundary value of the field $A_T$, we then
immediately have that $A_T=\overline{C}+\sigma \overline{D}$. A
simple calculation then reveals that
\begin{equation}
A_x'(u=0)=2\sigma A_T\left[1+\frac{\gamma}{2} \left[\frac{185}{4}+2\tilde{\alpha} \right]\right] \, .
\end{equation}
The on-shell action is given by
\be
S_{\textrm{total}} =- \frac{{\tilde {N}}^2 r_0^2}{16 \pi^2 R^4} \int
\frac{d^4k}{(2 \pi)^4} \int_0^1 du \, \left[ \frac{1}{2}A_{-k}
{\mathcal{L}} \, A_k +\,\partial_u \Psi \right]  \, .
\label{ActionOnShell}
\ee
where ${\mathcal{L}} A_k =0$ is simply the equation of motion,
and $\Psi$ is a boundary term. Upon evaluating
the on-shell action, the only surviving term is the boundary term,
as we expect from holography. This is given by
\ba
\Psi&=&(B_1+\gamma B_W-\gamma A_W)A_k'A_{-k}+\frac{\gamma}{2}(C_W-A_W')A_k A_{-k}-\gamma E_W'A_k''A_{-k}\nonumber \\
&&+ \gamma E_W A_k''A_{-k}'-\gamma E_W A_k''' A_{-k}
+\gamma E_W \left(p_1 A_k'+2 p_0 A_k\right) A_{-k}'-\gamma \frac{F_W'}{2} A_k' A_{-k} \, .\label{finalb}
\ea
The functions $A_W, C_W,D_W$ start at
${\cal{O}}(\varpi_0^2,\kappa_0^2)$, and so do not contribute to the
order of momentum in which we are interested. The function $E_W$
starts at ${\cal {O}}(u^2)$, and the regularity of $A_x$ at the
boundary ensures that the contribution from terms containing $E_W$
vanishes at the horizon. Therefore, we only get contributions from
$B_1$, $B_W$ and $F_W$ inside $\Psi$. Remembering that $r_0=\pi T
R^2(1-265/16 \gamma)$, we obtain that the conductivity is then
corrected by a factor
\be
1+\gamma \left(\alpha -10 \right) \, ,\nonumber
\ee
where
\begin{equation}
\alpha=\tilde{\alpha}-h_1-3 h_2 \, .
\end{equation}
%

\section{Conclusions}

In this work we have considered a strongly-coupled SCFT plasma at
finite temperature with ${\cal {O}}(\alpha'^3)$ corrections from
type IIB string theory using the gauge/string duality. The
corrections include those derived from the Ramond-Ramond five-form
field strength and consist of a set of dimension-eight operators in
the ten-dimensional type IIB supergravity action. We focused on the
effect of these dimension-eight operators on the behaviour of vector
fluctuations of the supergravity background. The
${\mathcal{O}}(\alpha'^3)$ corrections to type IIB supergravity are
dual to $1/{\lambda}^{3/2}$ corrections to the large $N$ limit of
the SCFT. Our aim was to study the vectorial fluctuations of the
ten-dimensional supergravity background in order to get an insight
into $R$-charge transport in the SCFT. In a certain limit
\cite{CaronHuot:2006te}, $R$-charge transport can be equated to EM
transport, essentially because a weak gauging of the $R$-symmetry
$U(1)$ allows us to embed $U(1)_{em}$ into the theory, and then
reinterpret our results with this is mind.

We recall that the type IIB string theory action with leading
${\mathcal{O}}(\alpha'^3)$ corrections is given by the usual
two-derivative minimal Lagrangian plus a number of eight-derivative
operators. The tensor structure of these operators is given by
contractions of four factors which are the Weyl tensor and the
${\mathcal{T}}$ tensor, where the latter is constructed from the
five-form field strength $F_5$ of type IIB string theory. This
complicated Lagrangian leads to a background which is a
{\emph{warped}} product of a deformed AdS-Schwarzschild black hole
with a compact five-dimensional Einstein manifold $M_5$, providing
that ${\cal{T}}_0$ vanishes. We emphasize that the corrections in
AdS factor of the background are independent of the internal
manifold as long as the latter is Einstein \cite{Buchel:2008ae}.
Using this corrected background we consider vector perturbations of
the metric and investigate the dimensional reduction of the
ten-dimensional type IIB string theory action at
${\mathcal{O}}(\alpha'^3)$ on the $M_5$. This leads to a
five-dimensional effective action for the $U(1)$ gauge fields, and
we study the full set of five-dimensional operators induced by the
ten-dimensional operators $Weyl^4+ Weyl^3 {\mathcal{T}}+Weyl^2
{\mathcal{T}}^2$. We find 26 independent five-dimensional operators
of the form $C^2 F^2$ and $C^2 (\nabla F)^2$, $F^2$ and $(\nabla
F)^2$. One of these operators vanishes on-shell leading to only 25
general five-dimensional scalar operators. In principle this
Lagrangian can be used to study the finite-coupling corrections to
the two-point functions of $R$-symmetry currents in a wide range of
strongly-coupled SCFTs.

As an application of our general effective Lagrangian, we then study the
transverse gauge fields $A_x$, whose solution we need in order to
obtain the conductivity of the QGP. In order to solve the EOM of
$A_x$ in the hydrodynamic regime we require ingoing boundary
conditions at the horizon. Importantly, we find that the frequency
of the waves at the horizon does not change compared with the
infinite 't Hooft coupling results. Our results indicate that any gauge
invariant Lagrangian for the vector perturbations yields an equation
of motion with the same singularity structure and indices at the
horizon. Finally, we obtain a general expression for the leading 't
Hooft coupling corrections to the conductivity for any Lagrangian
quadratic in the gauge field and containing up to four derivatives.

The results of this work constitute a step towards the understanding
of charge-transport at finite-coupling in a range of SCFTs. We
emphasize that the set of operators enumerated in here are present
in any type IIB string theory holographic dual model, because the
order ${\mathcal{O}}(\alpha'^3)$ ten-dimensional terms are present
regardless of the details of the dual theory. Thus, computing the
conductivity in this setting is of intrinsic value even if the exact
coefficients of the operators in five dimension are unknown. For
simple internal manifolds the exact contributions of the
ten-dimensional terms can be determined exactly. For $S^5$, where
the dual theory is ${\cal{N}}=4$ SYM, we will consider the full
ten-dimensional calculation of the conductivity with
$1/{\lambda}^{3/2}$ corrections in a future work \cite{ten-d}.

We end with a word on the universality of the corrections computed
here. We recall that Buchel {\emph{et al.}} found that the
corrections to the shear viscosity to entropy density ratio for all
theories dual to $AdS_5 \times M_5$, where $M_5$ is Einstein, was a
universal quantity. This was shown by proving that the
five-dimensional Lagrangian for the tensorial metric perturbations,
which govern both the viscosity and entropy of the theory, is the
same regardless of the internal manifold. This was obviously a very
exciting result, as it shows us a universal feature shared by the
strongly-coupled regime of a wide class of theories which have
totally different field contents in the weak-coupling description.
Naturally, we must ask if there is a possibility that the
five-dimensional effective Lagrangian we compute above also exhibits
universal behaviour. The problem we have here is that the operators
with coefficients $h_i$ have a direct dependence on the Weyl tensor
of the internal manifold, making it seem unlikely that the $h_i$ are
manifold-independent. However, the operators with coefficients $a_i
\longrightarrow g_i$ have no such dependence on the internal
manifold, and it could be that those coefficients are universal.
These are, however, very speculative comments, and clearly require a
lot of work to settle them.

~

\centerline{\large{\bf Acknowledgments}}

~

We thank Miguel Paulos, Kasper Peeters and Andrei Starinets for
useful discussions. B.H. thanks Christ Church College, Oxford, for
financial support, and also kind hospitality at CERN during Planck
2010 where part of this work has been carried out. The work of M.S.
has been partially supported by the CONICET and the ANPCyT-FONCyT
Grant PICT-2007-00849.

\newpage
\appendix

\section*{Appendix: The full Lagrangian}

The Lagrangian for the higher-dimensional terms is given by:
\ba\label{AW}
A_W&=&-2 u^5\left[a^w_1 f(u)\kappa_0^2+a^w_2\varpi_0^2 \right] +\tilde{A}_W\, , \nonumber \\
B_W&=&- 4 u^4 \left[(b^w_1-b^w_2 u^2+ b^w_3 u^4) -  b^w_5  \varpi_0^2 u
      - b^w_4 u f(u) \kappa_0^2 \right] +\tilde{B}_W\, , \nonumber \\
C_W&=&-4 \frac{u^4}{f(u)} \left[ 3 f(u)(c^w_1 u^2-c^w_2)  \kappa_0^2
      +\left(c^w_4-c^w_3 u^2\right) \varpi_0^2 \right] +\tilde{C}_W \, , \nonumber \\
D_W&=&-\frac{u^3}{f^2(u)} \left[d^w_6 u f^2(u) \kappa_0^4 +d^w_7 u
      \varpi_0^4 + d^w_8 u f(u)\kappa_0^2 \varpi_0^2 \right. \nonumber \\
&&\left. +4 f^2(u)(d^w_1 u^2-d^w_2)\kappa_0^2
      + 4 (d^w_3 -d^w_4 u^2 +d^w_5 u^4)\varpi_0^2\right]+\tilde{D}_W \, ,\nonumber \\
E_W&=& - e^w_1 u^6 f^2(u)+\tilde{E}_W  \, , \nonumber \\
F_W&=& 4 u^5 f(u) (f^w_1 u^2-f^w_2)+\tilde{F}_W \, ,
\ea
where we denote the contributions coming from $F^2$ and $\nabla F^2$
by $\tilde{A}_W$, $\tilde{B}_W$, $\tilde{C}_W$, $\tilde{D}_W$,
$\tilde{E}_W$, $\tilde{F}_W$, and we write the rest explicitly.
From our Lagrangian of Eq.(\ref{effective-L}) we have the identifications:
\ba
a^w_1&=& 2 \, (72 a_1 - 4 b_1 + 8 b_2 - 2 c_2 + 5 c_3 - c_4 + 2 c_6
         - c_7 + d_3 + d_4) \, , \nonumber \\
a^w_2&=& 2 \, (-72 a_1 + 4 b_1 - 12 b_2 + 6 c_2 + 3 c_3 + 9 c_4 - 2
c_6
+ 9 c_7 + 3 d_1 +  3 d_2 - 2 d_3 + d_4) \, , \nonumber \\
b^w_1&=& \frac{1}{4} \, \left( 2 [11 c_1 + 3 (13 c_3 + 2 c_4 + c_5 -
         2 (3 c_7 + c_8))] + 2 (-2 d_1 - 5 d_2 + 9 d_3 + 4 d_4)  \right. \nonumber \\
  &&  \left. - 36 (-6 a_2 - e_1) + 8 (16 b_3 + 22 b_4 + f_1) - 5 g_1 + 2 g_2 + g_3 \right) \, , \nonumber \\
b^w_2&=& \frac{1}{4} \, (576 a_2 - 32 b_2 + 352 b_3 + 480 b_4 + 64 c_1 + 32 c_2
       + 256 c_3 + 80 c_4+ 16 c_5 - 48 c_7 - 32 c_8 \, \nonumber \\
  &&  + 8 d_1 - 8 d_2 + 40 d_3 + 32 d_4 + 36 e_1 + 8 f_1 - 5 g_1 + 2 g_2 + g_3) \, , \nonumber \\
b^w_3&=& \frac{1}{4} \, (-576 a_1 + 504 a_2 + 32 b_1 - 128 b_2 + 320
b_3 + 432 b_4 + 62 c_1 + 80 c_2 + 238 c_3 + 132 c_4 \nonumber \\
  &&  + 14 c_5 - 16 c_6 + 20 c_7 - 28 c_8 + 28 d_1 +
   14 d_2 + 18 d_3 + 32 d_4) \, , \nonumber \\
b^w_4&=&-36 a_2 - 16 b_3 - 24 b_4 - c_1 - c_3 + 6 c_4 - c_5 + 2 (c_7 + c_8) + d_1 \, , \nonumber \\
b^w_5&=& 36 a_2 + 24 b_3 + 32 b_4 + 5 c_1 + 9 c_3 - 2 c_4 + c_5 - 10 c_7 - 2 c_8 -
 2 d_1 - 3 d_2 + 3 d_3 \, , \nonumber \\
c^w_1&=& \frac{2}{3} (-72 a_1 - 36 a_2 + 4 b_1 - 10 b_2 - 16 b_3 - 20 b_4 + c_1 + 4 c_2
       - 10 c_3 + 5 c_4 - c_5 - 2 c_6 + 3 c_7  \nonumber \\
       && + 2 c_8 + d_1 - 2 (d_3 + d_4))\, , \nonumber \\
c^w_2&=& - \frac{2}{3} \,  (36 a_2 + 2 b_2 + 16 b_3 + 20 b_4 - c_1 -
          2 c_2 + 5 c_3 - 4 c_4 + c_5 - 2 c_7 - 2 c_8 - d_1 + d_3 + d_4) \, , \nonumber \\
c^w_3&=& -144 a_1 + 36 a_2 + 8 b_1 - 28 b_2 + 24 b_3 + 32 b_4 + 5
c_1 + 16 c_2 + 23 c_3 + 24 c_4 + c_5 - 4 c_6 + 16 c_7 \nonumber \\
       && - 2 c_8 + 7 d_1 + 6 d_2 - 2 d_3 + 4 d_4 \, , \nonumber \\
c^w_4&=& -72 a_2 - 4 b_2 - 48 b_3 - 64 b_4 - 10 c_1 + 4 c_2 - 10 c_3
+ 12 c_4 - 2 c_5 + 28 c_7 + 4 c_8 + 7 d_1 + 9 d_2 \nonumber \\
 &&  - 7 d_3 + 2 d_4\, , \nonumber \\
d^w_1&=& \frac{1}{4} \, (-108 a_2 - 56 b_3 - 64 b_4 + c_1 - 3 c_3 + 22 c_4
- 3 c_5 + 6 c_7 + 6 c_8 + 4 d_1 - d_2 + d_3)\, , \nonumber \\
d^w_2&=& \frac{1}{4} \,  (-108 a_2 - 56 b_3 - 64 b_4 + c_1 - 3 c_3 + 22 c_4
- 3 c_5 + 6 c_7 + 6 c_8 + 4 d_1 - d_2 + d_3 - 36 e_1\nonumber \\
&&  - 4 f_1 - 3 g_1 - 2 g_2 - g_3) \, , \nonumber \\
d^w_3&=& \frac{1}{4} \,  (-108 a_2 - 64 b_3 - 104 b_4 - 11 c_1 - 19 c_3 + 10 c_4
- 7 c_5 + 26 c_7 + 14 c_8 + 8 d_1 + 7 d_2 - 7 d_3 \nonumber \\
&&   - 36 e_1 - 8 f_1 + 5 g_1 - 2 g_2 - g_3) \, , \nonumber \\
d^w_4&=& \frac{1}{4} \,  (16 b_3 - 16 b_4 + 8 c_1 + 16 c_3 + 8 c_4 - 8 c_5 - 8 c_7
+ 16 c_8 + 4 d_1 - 4 d_2 + 4 d_3 - 36 e_1 - 8 f_1 \nonumber \\
&&   + 5 g_1 - 2 g_2 - g_3)\, , \nonumber \\
d^w_5&=& \frac{1}{4} \,  (-36 a_2 - 16 b_3 - 40 b_4 - c_1 - c_3 + 6
c_4 - 5 c_5 + 6 c_7 +
 10 c_8 +  4 d_1 + d_2 - d_3) \, , \nonumber \\
d^w_6&=& -2 \, (72 a_1 - 36 a_2 - 4 b_1 + 4 b_2 - 8 b_3 - 16 b_4 + 3 c_1 + 2 c_2
- 4 c_3 -  7 c_4 - c_5 + 2 c_6 + c_7 \nonumber \\
&& + 2 c_8 - d_1 - d_3 - d_4) \, , \nonumber \\
d^w_7&=&2 \, (-72 a_1 + 36 a_2 + 4 b_1 - 12 b_2 + 24 b_3 + 32 b_4 + 5 c_1 + 6 c_2
+ 12 c_3 + 7 c_4 + c_5 - 2 c_6 - c_7\nonumber \\
&& - 2 c_8 + d_1 + d_3 + d_4) \, , \nonumber \\
d^w_8&=&4 \, (72 a_1 - 36 a_2 - 4 b_1 + 8 b_2 - 16 b_3 - 24 b_4 - c_1 - 2 c_2
+ 4 c_3 +  5 c_4 - c_5 + 2 c_6 + c_7 + 2 c_8 \nonumber \\
&& + d_1 + d_3 + d_4) \, , \nonumber \\
e^w_1&=& 2 \, (-72 a_1 + 36 a_2 + 4 b_1 - 12 b_2 + 24 b_3 + 32 b_4 + 5 c_1
+ 6 c_2 + 12 c_3 + 7 c_4 + c_5 - 2 c_6 \nonumber \\
&& - c_7 - 2 c_8 + d_1 + d_3 + d_4) \, , \nonumber \\
f^w_1&=& \frac{1}{2} \, (-288 a_1 + 180 a_2 + 16 b_1 - 56 b_2 + 120 b_3
+ 160 b_4 + 25 c_1 + 32 c_2 + 73 c_3 + 42 c_4\nonumber \\
&& + 5 c_5 - 8 c_6 + 2 c_7
- 10 c_8 + 8 d_1 + 3 d_2 +  5 d_3 + 8 d_4) \, ,  \nonumber \\
f^w_2&=&  \frac{1}{2} \,  (108 a_2 - 8 b_2 + 72 b_3 + 96 b_4 + 15
c_1 + 8 c_2 + 43 c_3 + 10 c_4 +  3 c_5 - 14 c_7 - 6 c_8 - 3 d_2
\nonumber \\
&& + 7 d_3 + 4 d_4) \, . \nonumber
\ea
We also have the contributions given by
\ba\label{AWtilde}
\tilde{A}_W&=&0 \, , \nonumber \\
\tilde{B}_W&=&-h_1 f(u) + 2 h_2 (-3 + u (8 u - 7 u^3 - 2 \kappa_0^2 f(u)+ 2 \varpi_0^2))  \, , \nonumber \\
\tilde{C}_W&=&-8 h_2  \kappa_0^2 f(u) + \frac{4 h_2 (2 + u^2) \varpi_0^2}{f(u)}   \, , \nonumber \\
\tilde{D}_W&=&\frac{1}{u f(u)^2}\left(-\kappa_0^2 f(u)^2 (h_1 + h_2 (3 f(u)+ 2 \kappa_0^2 u )) \right. \nonumber \\
&&\left.+ (h_1 f(u) +
    h_2 (3 + u^4 +4 \kappa_0^2 u f(u))) \varpi_0^2 -
 2 h_2 u \varpi_0^4\right)  \, ,\nonumber \\
\tilde{E}_W&=&-2 h_2 u^2 f(u)^2  \, , \nonumber \\
\tilde{F}_W&=& -2 h_2 u f(u)(3 -  5 u^2)  \, .
\ea
%
%
%
%
%
%
%
%
%

\end{document}